# Light-induced quantum tunnelling current in graphene

Mohamed Sennary[1], Jalil Shah[1], Mingrui Yuan[1,2], Ahmed Mahjoub[3], Vladimir Pervak[4], Nikolay V. Golubev[1] and Mohammed Th. Hassan[1,2]*.

[1] Department of Physics, University of Arizona, Tucson, AZ 85721, USA.

[2] James C. Wyant College of Optical Sciences, University of Arizona, Tucson, Arizona 85721, USA

[3] Jet Propulsion Laboratory, California Institute of Technology, Pasadena, CA 91109, USA.

[4] Ludwig-Maximilians-Universität München, Am Coulombwall 1, 85748, Garching, Germany.

*Correspondence to: mohammedhassan@arizona.edu

**Abstract**

In the last decade, advancements in attosecond spectroscopy have allowed us to study electron motion dynamics in condensed matter. The access to these electron dynamics and, consequently, its control by an ultrafast light field paves the way for establishing ultrafast optoelectronics. Here, we report the generation of light-induced quantum tunnelling current in graphene phototransistors by ultrafast laser pulses in an ambient environment. This tunnelling effect provides access to the instantaneous field-driven current to demonstrate the current switching (ON and OFF) on a 630 attosecond (~1.6 petahertz speed). Moreover, we controlled the tunnelling current and enhanced the graphene phototransistor conductivity by controlling the density of the photoexcited charge carriers at different pump laser powers. Furthermore, we exploited this capability to demonstrate various logic gates. The demonstrated light-induced tunnelling current and ultrafast switching were attained under standard room temperature and pressure conditions. Therefore, the presented scientific advancement in this work is at the technology readiness level suitable for its immediate integration into the development of ultrafast—nearly six orders of magnitude faster—optical transistors, lightwave electronics, and optical quantum computers.

The development of ultrafast light tools is vital for studying light-matter interactions and related electron motion dynamics in real time[1-3]. For instance, the generation of XUV attosecond pulses via high-harmonic generation in the solid-state[4-7] permitted probing the strong field-induced electron dynamics in condensed matter[8-14]. Recently, the generation of a single attosecond electron pulse and the development of attomicroscopy have given access to the bound electron dynamics in the nanostructure and connected it to its morphology[15]. Moreover, the ability to manipulate and synthesis the waveform of ultrashort laser pulses allows for controlling the electronic motion, electronic structure, and physical properties of dielectric and semiconductor materials to demonstrate ultrafast optical switches[16-27]. Furthermore, both optical and XUV pulses have been used to generate ultrafast current signals[27-33]. These studies have found many applications, such as the demonstration of optical-based devices for sampling the ultrafast waveforms of light[33-52].

Recently, the generation of light-induced current ($I_L$), based on photoexcitation of graphene's carriers, has been reported[52-55]. This average $I_L$ is measured and demonstrated based on the flow of the excited carriers between two metal electrodes in a circuit. The $I_L$ has been controlled by manipulating the carrier dynamics by changing the intensity and the carrier-envelope phase (CEP) of the pump laser pulse[53,54]. It is noteworthy to mention that the $I_L$ current has a contribution from two currents: (i) The ultrafast instantaneous *field-induced current ($I_E$)*, which is generated from the motion of the excited virtual carriers—driven by the light field—in the conduction band (intraband current) of graphene. This $I_E$ current is a transient current and lasts only during the laser pulse field time window. (ii) The *photo-induced current ($I_p$)*, which is generated due to the excitation of real carriers from the valance band to the conduction band by absorbing photon(s) from the pump pulse (interband current). Then, these excited carriers relax back to the valance band on a time scale of a few ten picoseconds. Hence, the major contribution in $I_L$ is coming from $I_p$, while the contribution of $I_E$ is minor since the latter exists in a finite time (the duration of the laser pulse in femtosecond time scale) compared to the long-time response of the current detector (in few milliseconds time scale).

In previous studies using symmetric graphene[52-55], the *$I_E$* was averaging out, and the measured current was mainly from the *$I_p$* contribution. Additionally, the demonstrated control by changing the CEP is based on the modulation of the excited real carrier's density and the displacement of virtual carriers in real space by changing the pump pulse intensity[53]. Nevertheless, the detection and distinguishing of the *$I_E$* haven't been measured or demonstrated yet.

In this work, we utilised a graphene-silicon-graphene (Gr-Si-Gr) phototransistor to generate sub-microamperes light-induced current ($I_L$) by few-cycle laser pulses. In our transistor, the current flows based on quantum tunnelling between the graphene sides through the silicon junction. Hence, the generated current is gated in time, which allows us to access and record the ultrafast instantaneous field-induced current ($I_E$). The $I_E$ modulates periodically in real-time, following the waveform of the driver field, enabling a current switching between two states (ON and OFF) with a time speed of 630 attoseconds (1.6 petahertz). Moreover, we control the $I_L$ current amplitude by increasing the induction laser beam intensity and determine the consequent enhancement of our phototransistor photoconductivity. Finally, the flexibility of our transistor setup allowed us to combine a DC current ($I_V$), generated by applying external voltage, with the $I_L$, to demonstrate several logic gates within our phototransistor. Importantly, the presented experiments are performed under ambient standard temperature and pressure conditions, making this phototransistor at the technology readiness level for developing attosecond and lightwave quantum optoelectronics.

**Light-induced quantum tunnelling current and attosecond current switching**

The development of graphene field-effect phototransistor based on quantum tunnelling is essential to access the field-induced current in graphene. Hence, we optically dopped a graphene-based channel transistor to prepare a Gr-Si-Gr channel (the preparation and the operation mechanism of our device are explained in SI and illustrated in Fig. S1). Optical microscope images of this channel and an illustration of its band structure are displayed in Fig. 1a. The Gr-Si-Gr composition is confirmed by Raman spectroscopy characterisation measurements and results (as explained in SI and shown in Fig. S2). Initially, we maintained the external voltage ($V_{ext}$) in our device at zero voltage, ensuring that no external DC current is generated ($I_V=0$). Then, we focused ultrafast laser pulses (the measured temporal profile (FWHM~6.5 fs) is shown in Fig. S3a) by a parabolic mirror into the Gr-Si-Gr channel (see details in Methods, SI, and Fig. S2b). Hence, a light-induced current signal ($I_L$) —in the few hundred nanoamperes level—is generated and measured (see Fig. 1b). Note this $I_L$ signal switches OFF when the Laser beam is blocked, as shown in see Fig. 1b. When the laser is ON, the graphene charge carriers are excited, leading to an increase in their concentration. Hence, the density of states changes, causing the Fermi energy level to shift from the neutral level and increasing the voltage difference, as demonstrated elsewhere[56] (see illustration on the right side of Fig. 1a). Accordingly, the generated $I_L$ flows in our device by quantum tunnelling of the carriers between the two graphene sides. To prove this current tunnelling, we measured the IV curves in both cases (laser ON and OFF), as shown in Extended Data 1. From these measurements, we obtained the IV curve shown in Fig. 1c by subtracting the IV curve when the laser is OFF from the IV curve when the laser is ON (after shifting it to compensate for the $I_L$ offset). Remarkably, this curve (Fig. 1 c) is a tunnelling characteristics IV curve, validating our interpretation of the $I_L$ generation and flow mechanism[56].

The current tunnelling flow mechanism gates the generated current signal in time and allows us to measure and distinguish the instantaneous field-induced current ($I_E$), which is generated due to the intraband dynamics in graphene. This current evolves during the laser pulse's existence time window. Hence, to measure $I_E$ in real-time, we opted to perform a cross-correlation current measurement between two current signals generated by two pump laser pulses. Accordingly, we modified our setup by splitting the input laser beam into two beams using a beamsplitter (see Methods and Fig. S.3c); each beam power has been set to have a similar estimated field strength of ~0.85 V/nm. Then, we recorded the current as a function of the time delay between the two pulses. Our setup's capability enabled the compensation of the $I_P$ current (generated by the interband dynamics) by applying an external voltage until the output measured current is zero amperes when the two laser pulses are not overlapping in time. The average of three cross-correlation current measurements is shown in Fig. 2a (black dots connected with red lines). A minor contribution of the current amplitude oscillation (Fig. 2a) potentially originated from the optical interference, since we observed only a 10% oscillation in the power between the two pulses at the temporal overlap. Please note the two beams aren't collinearly propagating, and they incident on the sample with small angles (< 5°) (Fig. S2c), which minimises the optical interference effect. Furthermore, the absolute measured $I_E$ current amplitude signal in real-time (plotted in Fig. 2b) switches from 29 nA (ON status) to <1 nA (OFF status) in 630 attoseconds (see the inset of Fig. 2b), demonstrating the attosecond current switching in our phototransistor.

We attributed this measured current oscillation to the drifting of the excited carriers within the conduction band (intraband current) of graphene following the driver laser field[27,34,53,57,58]. To confirm our observations, we performed quantum mechanical calculations to simulate our experiment current measurements. In our calculation, we first assumed that the measured cross-correlation current (Fig. 2a) reflects the cross-correlation of the laser fields. Hence, we decomposed the waveform of the driver pulse (plotted in the red line in Extended Data 2) from the cross-correlation profile in Fig. 2a. Notably, the temporal profile of the deconvoluted waveform and the measured temporal profile of the pump pulse (Fig. S2a) are in a good agreement, validating our assumption. We utilised this waveform in our quantum simulation model after considering the tunnelling effect by adding a complex absorbing potential (CAP), as explained in Methods. Then, we calculated the generated net current after the action of the two pulses as a function of the time delay (Extended Data 3) and plotted it (dashed black line in Fig. 2a) in contrast with the measured current. These two currents are in good agreement and follow the pump pulse waveform. Noteworthy, when we ignore the tunnelling effect (CAP) in our calculations, the calculated cross-correlation net current is zero (plotted in Extended Data 4a &b) due to the averaging out of the $I_E$ current, which confirms the pivot effect of the tunnelling in our $I_E$ measurements and explains why the previous studies were not able to access or measure this field-driven current [53,54].

**Controlling the light-induced tunnelling current and photoconductivity in graphene.**

The photo-induced $I_P$ current signal has the main contribution to the $I_L$ current. $I_P$ is generated from the interband current dynamics in graphene. Hence, the amplitude of $I_P$ depends on the excited charge carrier's density and its distribution in the reciprocal space, which can be controlled by changing the intensity of the exciting pulse. Thus, we measured the $I_L$ amplitude at different field intensities of the pump laser ranging from 0 to 2 V/nm and plot the result in Fig. 3a (black dots connected by red line). The $I_L$ amplitude increases gradually and then reaches a plateau at a higher intensity. This can be attributed to the increase in the number of excited carriers and the population in the conduction band before the carriers reach saturation. Accordingly, we calculated the average excited carrier population at different field intensities by solving the time-dependent Schrödinger equation (more details are provided in Methods) and plotted it in the blue line in Fig. 3a. The calculated carrier population exhibits dynamic behaviour and plateau similar to the measured current $I_L$ shown in Fig. 3a. Moreover, Fig. 3b shows the distribution of excited carriers pumped by ~1.5 V/nm in the reciprocal space of graphene. In Fig. 3b, the ring structure around the Dirac point ($k_x = k_y = 0$) reflects the single-photon excitation region. The presence of the population in the vicinity of the Dirac point is due to the temperature effects (see Methods), which is considered in our calculations. Also, Fig. 3b shows a very minor exciting start to appear from two-photon absorption. These results explain the linear trend and the plateau in the measured current (Fig. 3a) as a saturation of the single-photon excitations. Increasing the field intensity even further is expected to show a nonlinear behaviour increase in the current due to the increase in the two-photon excitation contribution, which, however, cannot be observed in our measurement since we observed a damaging effect at higher field strength.

Next, we studied the effect of the light-induced current and carrier excitation on the resistivity and photoconductivity of our phototransistor[59]. Thus, we measured the IV curves at different pump laser field intensities. The results are shown in Fig. 4a. The asymmetry in the positive and negative voltage sides is due to the generation of $I_L$ with different values as the intensity increases. We focused our measurement on the intensity range before the saturation (from 0-1.2 V/nm). From the slopes of the measured IV curves (in Fig. 4a), we calculated the resistance (R) as a function of the field intensity. The resistance of the phototransistor remains the same until a certain intensity, then it decreases from ~6 to less than 5.6 KΩ at 1.2 V/nm, as shown in Fig. 4b. Accordingly, the phototransistor conductivity increased by ~7.5% (Fig. 3c, black points). The blue line in Fig. 3c shows the simulation fitting of the conductivity change at different intensities (as explained in the Methods).

The controlling of the $I_L$ signal (hereafter referred to as signal A) and the DC current ($I_V$) (referred to as signal B) by adjusting laser pulse intensity and the applied external voltage in our phototransistor, respectively, allow us to demonstrate various optical logic-gates. For instance, by applying $V_{ext}$ of -3.6 mV, we generated $I_V$ current of 600 nA; this effectively cancelled the induced $I_L$ current (-600 nA, shown in the inset of Fig. 1b). Consequently, our device measures no output current signal, demonstrating the XOR & NOT logic gates (see Table 1 & 2). When adjusting the applied $V_{ext}$ such that the $I_V$ is < $I_L$, the output current signal ≠ 0. In this case, we can establish the logic gate OR, as shown in Table 3. Moreover, by exploiting and illuminating all seven single-graphene channels and the seven triple-graphene transistor channels in our device by different power-controlled laser beams simultaneously, we can create a multichannel phototransistor (operating with laser repetition rate) and establish all possible logic gates for developing digital quantum tunnelling-based photonics devices.

**Conclusion**

In this work, we demonstrate the light-induced quantum tunnelling current in a Gr-Si-Gr phototransistor. The current flow is based on the tunnelling of electrons between the graphene terminals through the Si Junction. This current has more than three orders of magnitude better efficiency than the typical graphene transistor[53,54]. Moreover, this high efficiency led to generating a decent light-induced current amplitude at low pumping laser power. Hence, this Gr-Si-Gr transistor can operate in ambient conditions (normal pressure and temperature conditions) in analogy to the typical graphene phototransistor, which operates in vacuum to avoid the oxidation of graphene and the degradation of the transistor when illuminated with a high-intense laser beam. Furthermore, the presented current tunnelling mechanism in the Gr-Si-Gr transistor gate the laser field-induced current signal; thus, it subsists after the pulse, which is not possible in a symmetric graphene transistor. Hence, this ultrafast current—which has a sub-femtosecond switching time—can be logged, demonstrating the petahertz current switching speed in our transistor. Furthermore, the tunnelling effect led to dynamic modification of the resistivity and conductivity of the phototransistor. We report a reduction in the transistor photoresistivity by ~0.4 KΩ, which corresponds to an enhancement of 7.5% in the photoconductivity. Hence, this work promises to advance the scientific and technological advancements of ultrafast lightwave quantum electronics, attosecond optical switches, and ultrafast data encoding and communication[18,26]. Moreover, the ability to optically control the light-induced quantum current signal and establish different optical logic gates open the door for developing ultrafast quantum optical computers.

**Methods**

**Experiment setup**

In our setup, a 1 mJ few-cycle laser pulse centred at 750 nm is generated from an OPCPA-based (passively carrier-envelope phase (CEP) stabilised) laser system with a 20 kHz repetition rate. A supercontinuum laser beam that spans over 400-1000 nm is generated by focusing the laser beam in a hollow-core fibre (HCF). This supercontinuum enters a chirp mirror compressor to generate a ~6.5-fs laser pulse. The measured temporal profile using the FROG technique is shown in Fig. S3a. The laser beam is focused (beam diameter is ~50 µm) on one of the transistor channels by using a 25 mm parabolic mirror Fig. S3b. The graphene chip is connected to an external voltage and current source/detector. This device is used to measure the light-induced current signal $I_L$ (see SI). To measure the field-induced current $I_E$, the output beam from the chirp mirror compressor splits into two beams by beamsplitter. One of the beams reflects off two mirrors mounted on a delay stage (piezo stage) with nanometer resolution and is combined with the second beam by another beamsplitter (Fig. S3c). Then, the two beams are sent to the same parabolic mirror and focus into the graphene chip. The $I_E$ is recorded as a function of the time delay between the two pulses.

**Simulations of the excited carrier dynamics and generated currents in graphene**

The light-induced population transfer dynamics in graphene can be obtained by solving the semiconductor Bloch equation[52,54,57]:

$$i\hbar \frac{\partial}{\partial t} \rho_{m,n}(\boldsymbol{k}, t) = [E_m(\boldsymbol{k}_t) - E_n(\boldsymbol{k}_t)]\rho_{m,n}(\boldsymbol{k}, t)$$

$$+\boldsymbol{E}(t) \cdot \{\boldsymbol{D}(\boldsymbol{k}_t), \rho(\boldsymbol{k}, t)\}_{m,n} - i\frac{1-\delta_{m,n}}{T_d} - W(\boldsymbol{k}_t)\delta_{m=n,C}\rho_{C,C}(\boldsymbol{k}, t) \quad (1)$$

where $\rho_{m,n}(\boldsymbol{k}, t)$ denotes the matrix element of the density matrix $\boldsymbol{\rho}(\boldsymbol{k}, t)$, the commutator symbol "{}" is defined as $\{A, B\} = AB - BA$, $T_d$ is the interband dephasing time, and the meaning of the term $W(\boldsymbol{k})$ will be explained later in this section. The electronic energies of the bands $E_i(\boldsymbol{k})$ and the corresponding matrix of the transition dipole moments $\boldsymbol{D}(\boldsymbol{k})$ are obtained by employing the tight-binding approximation. Eq. (1) is derived assuming the validity of the dipole approximation and using the Houston basis in the velocity gauge with the crystal momentum frame evolving according to the Bloch acceleration theorem:

$$\boldsymbol{k}_t = \boldsymbol{k} + \frac{e}{\hbar}\boldsymbol{A}(t), \qquad (2)$$

where $e$ is the elementary charge and $\boldsymbol{A}(t) = -\int_{-\infty}^{t} \boldsymbol{E}(t')\,dt'$ is the vector potential of the corresponding applied electric field $\boldsymbol{E}(t)$.

We simulated the temporal evolution of the density matrix $\boldsymbol{\rho}(\boldsymbol{k}, t)$ in reciprocal space by numerically solving Eq. (1). We sampled the unit cell in the first Brillouin zone with a uniform 256 × 256 grid along the reciprocal lattice vectors. The initial electron density was generated by employing the Fermi-Dirac distribution.

$$\rho_{n,n}(\boldsymbol{k}, t = 0) = \frac{1}{\exp{(E_n(\boldsymbol{k})/k_B T)}+1} \qquad (3)$$

where $k_B$ is the Boltzmann constant and the temperature $T$ is set to 298.15 K. The integration in the time domain is performed by the Runge–Kutta–Fehlberg method with adaptive time step control. We used a Gaussian waveform as follows:

$$E(t) = E_0 e^{-4ln2\left(\frac{t-t_0}{FWHM}\right)^2} \cos(\omega(t - t_0)) \qquad (4)$$

with a photon energy ω of ~1.5 eV linearly polarised along the C–C bonds of the graphene sample for modelling the applied electric field. The dephasing time $T_d$ is set to 10 fs.

The redistribution of the electron density between the valence and conduction bands of graphene under the influence of a time-dependent electric field affects the macroscopic properties of the material, such as the electrical conductivity. The latter per unit of volume can be obtained from the Kubo-Greenwood formula[60-62]:

$$\sigma_{\mu\nu} = \frac{e^2}{i\hbar} \sum_n \int_{BZ} \frac{\partial \rho_n(\varepsilon)}{\partial \varepsilon}\bigg|_{\varepsilon=E_n} \frac{\partial_{k_\mu} E_n(\boldsymbol{k}) \partial_{k_\nu} E_n(\boldsymbol{k})}{\hbar\omega_0 + i\eta}, \qquad (5)$$

where the $\mu$ and $\nu$ indices denote the directions of the conductivity tensor, $\sigma$. $\omega_0$ is the frequency of the applied, in general AC, spatially homogeneous test current. The infinitesimal imaginary shift $\eta$ added to the frequency acts as a small inelastic scattering rate or relaxation rate, and the integration is performed over the entire Brillouin zone. The derivatives of the energy distribution function for the valence and conduction bands are obtained from the corresponding residual population distributions, $\rho_V(\mathbf{k})$ and $\rho_C(\mathbf{k})$, respectively, using the following relationship connecting the corresponding partial derivatives relationship:

$$\frac{\partial \rho_n(\mathbf{k})}{\partial \mathbf{k}} = \left.\frac{\partial \rho_n(\varepsilon)}{\partial \varepsilon}\right|_{\varepsilon=E_n} \frac{\partial E_n(\mathbf{k})}{\partial \mathbf{k}}. \tag{6}$$

We computed the change in the electrical conductivity of the graphene sample as a function of the intensity of the applied electric field (see Fig. 4c). We used the DC test field in our simulations ($\omega_0 = 0$) and assumed the electron relaxation rate $\eta$ to be 0.01.

In addition to the residual change in the conductivity resulting from the action of the laser pulse on the system, the instantaneous intraband current generated during the action of the field can be estimated as follows:

$$\mathbf{J}^{intra}(t) = \sum_n \int_{BZ} \rho_{n,n}(\mathbf{k},t) \nabla_\mathbf{k} E_n(\mathbf{k}_t) d\mathbf{k}. \tag{7}$$

The application of a single isolated laser pulse to the graphene system generates an electron current, which, however, vanishes after the action of the laser field on the system is over. Similarly, applying two delayed laser pulses to symmetric graphene and performing the scan over various delays will not generate any residual current in the system, as shown in Extended Data 4.

However, in our case, the presence of the graphene-silicon junction breaks the symmetry of the system and potentially leads to electron tunnelling from the conduction band of graphene to silicon. To simulate the experimental measurements and to theoretically confirm that the field-induced current shown in Fig. 3a can exist, we added the possibility for electrons to tunnel from the conduction band of graphene to the silicon due to the created junction (see Fig. 1a and discussion in the main text). We added the complex absorbing potential (CAP) to Eq. (1) (the $W(\mathbf{k})$ term); this is a simple phenomenological way to account for electron leakage through a junction.

In our simulation, the CAP is chosen to be located to the right with respect to the Dirac point such that its strength increases along the $k_x$ direction of the crystal momentum $\mathbf{k}$:

$$W(\mathbf{k}) = \beta\theta(k_x - K_x)A_x(t)(k_x - K_x)^2, \tag{8}$$

where $K_x = \frac{2\pi}{\sqrt{3}a}$ is the coordinate of the Dirac point along the $k_x$ direction, $\theta(k_x - K_x)$ is the Heaviside step function, $A_x(t)$ is the $k_x$-component of the vector potential $\mathbf{A}(t)$, and $\beta$, chosen to be 5.0 in our simulations, is the parameter controlling the strength of the CAP. In the presence of CAP, the electron density can leak from the conduction band of graphene when it is displaced by the vector potential in the positive $k_x$ direction, or the electron density, in principle, can be pulled to the system from the junction when the vector potential is negative. The presence of the graphene-silicon junction breaks the symmetry of the system and thus leads to the generation of a persistent current after the interaction with the applied laser field, as shown in the results plotted in Extended Data 3b.

62    Huhtinen, K.-E. & Törmä, P. Conductivity in flat bands from the Kubo-Greenwood formula. *Physical Review B* **108**, 155108, (2023).**Competing interests**

The authors declare no competing interests.

**Data availability**

The datasets generated and/or analysed during this study are available from the corresponding authors upon reasonable request.

**Code availability**

The analysis codes that support the study's findings are available from the corresponding authors upon reasonable request.

**Figures and figure captions**

**Figure 1**

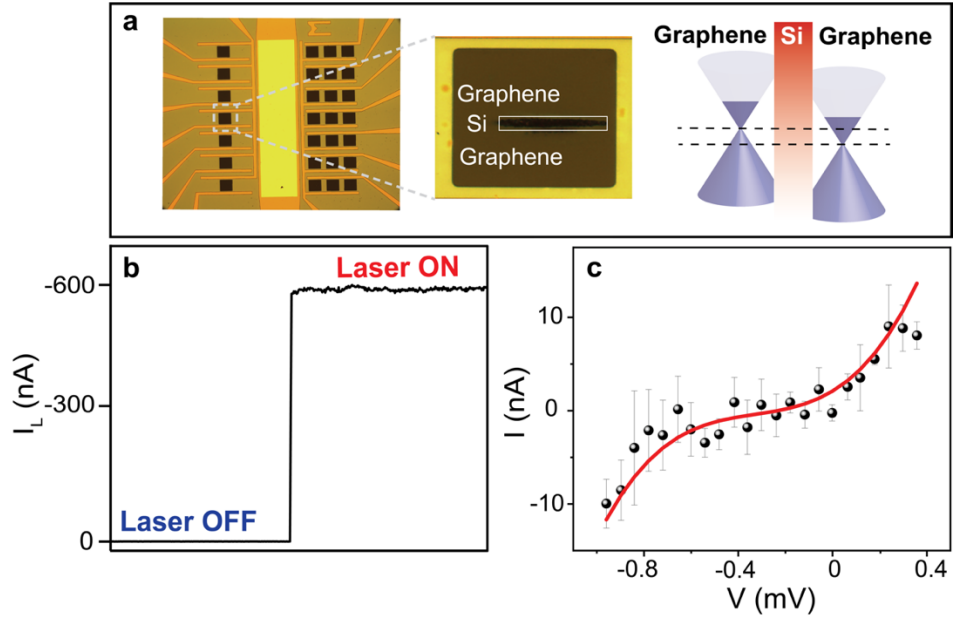

**Fig. 1. Light-induced quantum current tunnelling in graphene phototransistor. a** The optical microscope (and zoom in) images of the graphene-silicon-graphene phototransistor and illustration of its band structure, the black dashed line present the Fermi level. **b**, The light-induced tunnelling current $I_L$ generated when the laser beam is ON. **c**, The tunnelling characteristics IV curve for the Gr-Si-Dr transistor.

**Figure 2**

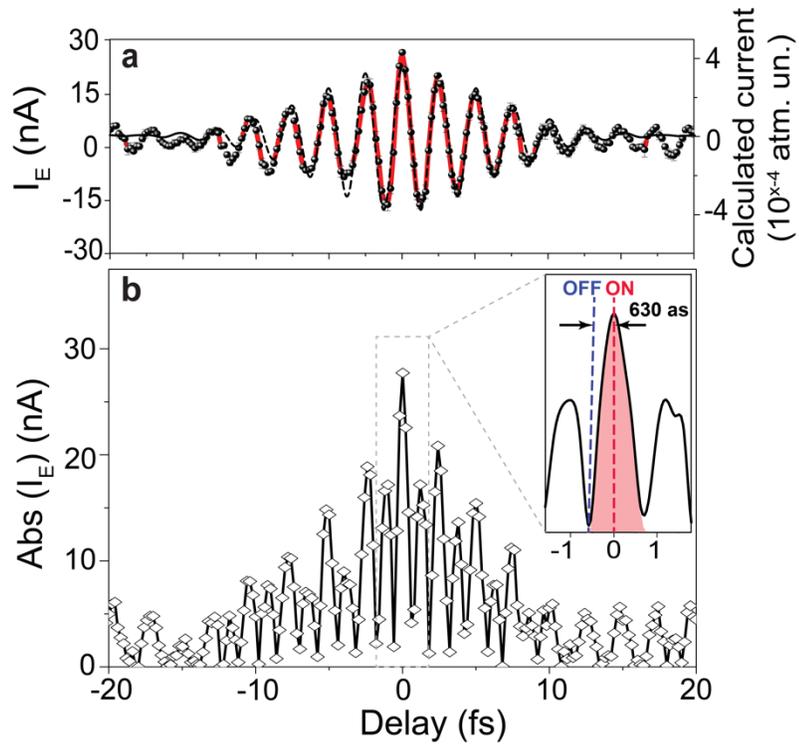

**Fig. 2. Field-induced current and attosecond current switching. a,** Instantaneous field-induced current ($I_E$) (average of three measurements), shown as black dots connected by red line. The calculated current is plotted in dashed black line. **b**, Absolute $I_E$ measured signal modulation in time. The inset shows the switching of the current ON and OFF with a periodicity of 630 as.

**Figure 3**

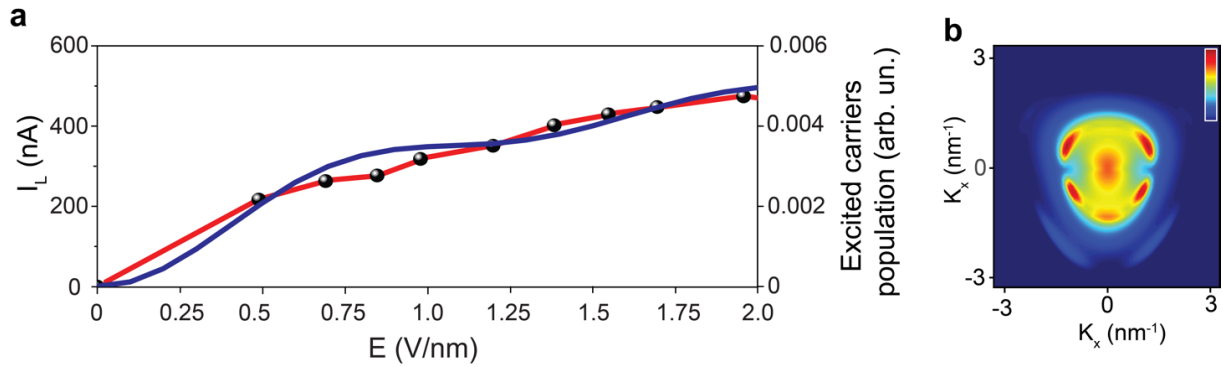

**Fig. 3. Controlling the light-induced $I_L$ current signal**. **a**, the measured light-induced current $I_L$ (black dotes connect by red line) and the calculated excited carriers' populations (blue line) as a function of the pump laser field intensity. **b,** Calculated carrier distribution in the reciprocal space of graphene excited by 1.5 V/nm laser field.

**Figure 4**

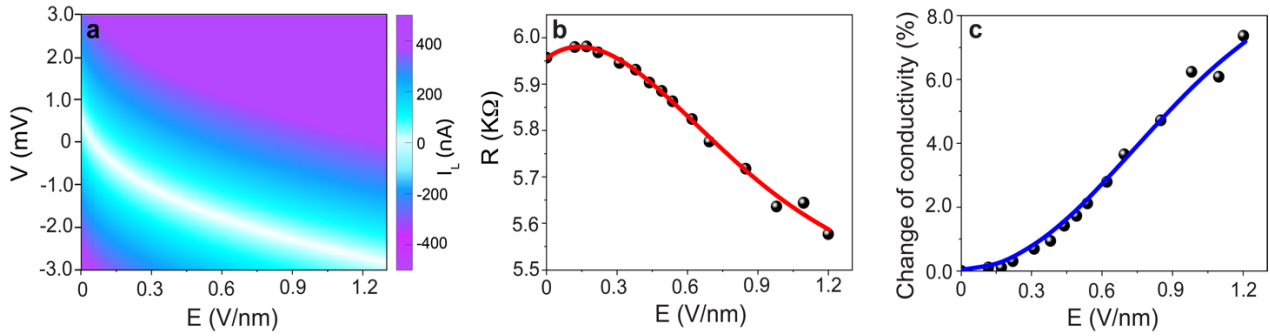

**Fig. 4. Photoconductivity enhancement in the graphene phototransistor. a,** Acquired IV curves (average of three measurements) at different pump field intensities. **b & c,** The change of resistance and conductivity as function of the laser field intensities obtained from the measured I-V curves in **a**, respectively. The blue line in **c** represents the calculated fitting conductivity from our simulation model.

# Tables

## Table 1

| Signal A: $I_L$ | | Signal B: $I_V$ | | Signal A XOR B: Measured output current | |
|---|---|---|---|---|---|
| OFF | 0 | OFF | 0 | OFF | 0 |
| ON | 1 | ON | 1 | OFF | 0 |
| ON | 1 | OFF | 0 | ON | 1 |
| OFF | 0 | ON | 1 | ON | 1 |

**Table 1. Demonstration of the XOR logic gate**. Signal A is $I_L$, where ON status means the laser beam is illuminating the Gr-Si-Gr transistor and OFF status means no laser is sent to the transistor. Signal B is the applied DC current $I_V$ by the external voltage source. ON and OFF status means sending -3.6 and 0 mV signal from the source.

## Table 2

| Signal A: $I_L$ | | NOT A: Measured output current | |
|---|---|---|---|
| ON | 1 | OFF | 0 |
| OFF | 0 | ON | 1 |

**Table 2. Demonstration of the NOT logic gate**. We adjust the external voltage to compensate for the generated $I_L$. In this case, no $I_L$ current is generated if the laser is ON; thus, the measured output signal will be OFF. Simultaneously, the measured current will turn to be ON, when the laser signal is switched off.

**Table 3**

| Signal A: $I_L$ | | Signal B: $I_V$ | | Signal A OR B: Measured output current | |
|---|---|---|---|---|---|
| OFF | 0 | OFF | 0 | OFF | 0 |
| ON | 1 | ON | 1 | ON | 1 |
| ON | 1 | OFF | 0 | ON | 1 |
| OFF | 0 | ON | 1 | ON | 1 |

**Table 3. Demonstration of the OR logic gate**. $I_V > I_L$, so the output current signal will be always ON unless both $I_V$ and $I_L = 0$.

**Extended Data**

**Extended Data 1**

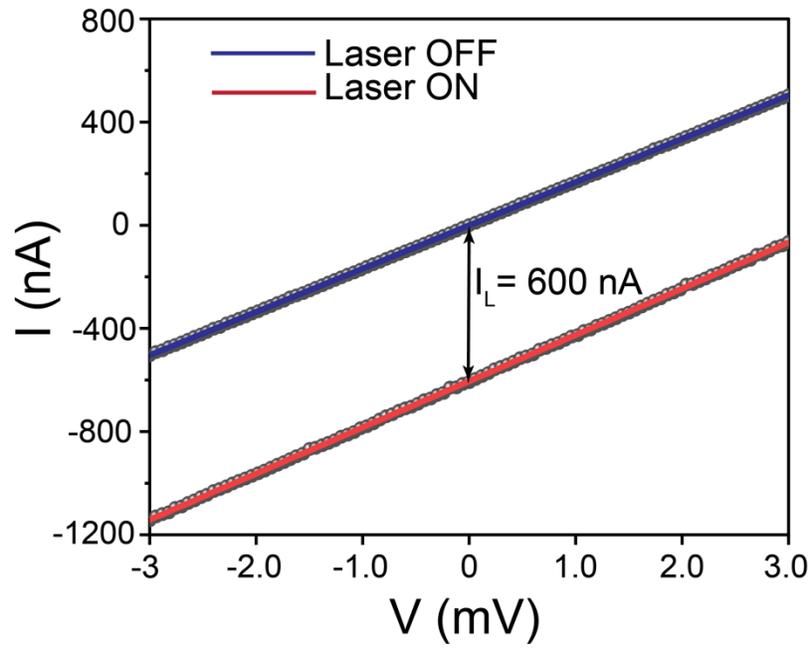

**Extended Data 1**. The measured IV curve in case of laser OFF (blue line) and laser ON (red line). The offset current presents the light-induced current ($I_L$) amplitude (600 nA).

**Extended Data 2**

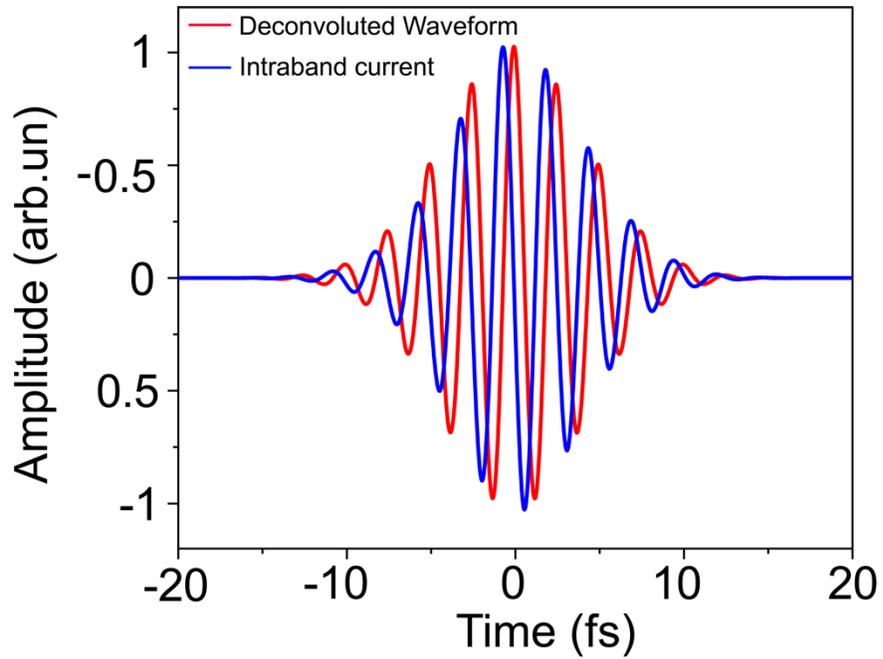

**Extended Data 2**. The retrieved deconvoluted waveform of the driver pulse from decomposition of the cross- corelation current measurement is plotted in red line. The intraband current simulated using fully quantum tight-binding model of graphene is depicted in blue line.

**Extended Data 3**

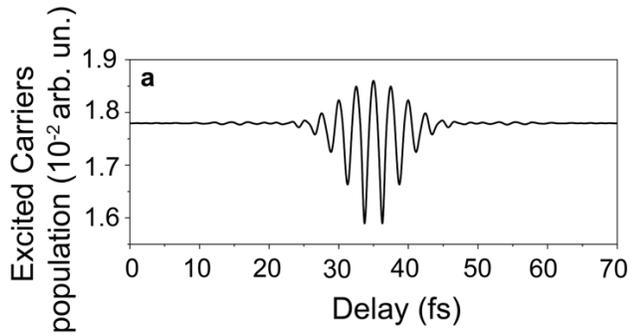 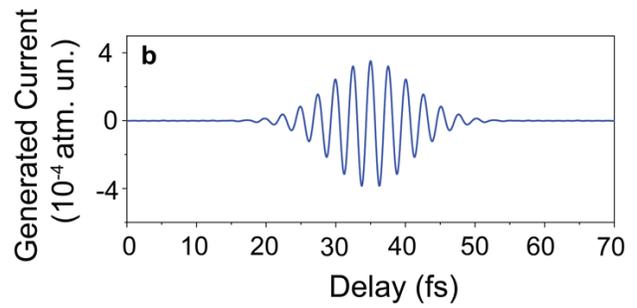

**Extended Data 3**. **a**, Calculated excited carrier's population, and **b**, the residual field-induced generated current as a function of the time delay between the two laser pulses in graphene junction.

**Extended Data 4**

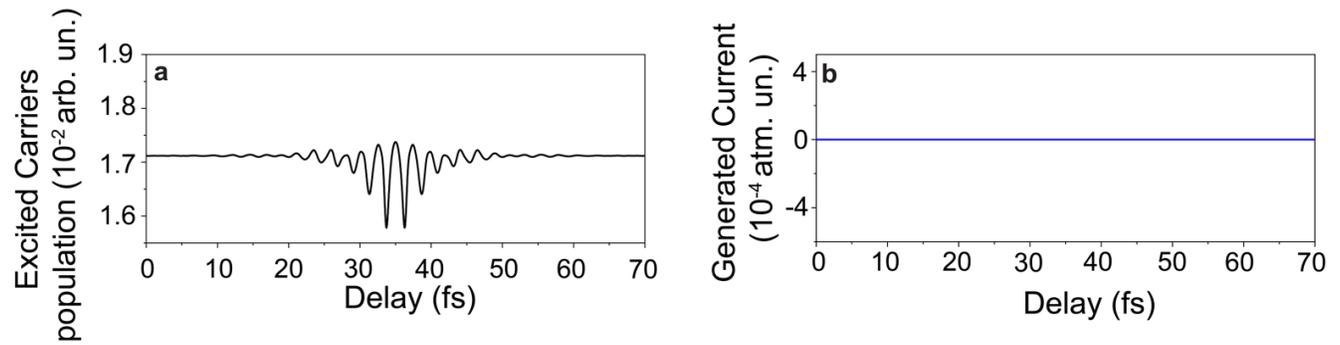

**Extended Data 4**. **a**, Calculated excited carrier's population, and **b**, the residual field-induced generated current as a function of the time delay between the two laser pulses in symmetric graphene.